\documentstyle[preprint,prb,aps]{revtex}
\newcommand{\be}{\begin{eqnarray}}
\newcommand{\ee}{\end{eqnarray}}

\begin{document}
\draft
\title{Optical Investigations of La$_{{\rm 7/8}}$Sr$_{{\rm 1/8}}$MnO$_3$}
\author{J. H. Jung,\cite{email1} K. H. Kim, H. J. Lee, J. S. Ahn, N. J. Hur, and T.
W. Noh\cite{lg}}
\address{Department of Physics, Seoul National University, Seoul 151-742, Korea}
\author{M. S. Kim and J.-G. Park}
\address{Department of Physics, Inha University, Inchon 402-751, Korea}
\date{\today }
\maketitle

\begin{abstract}
We investigated temperature dependent optical conductivity spectra of La$_{%
{\rm 7/8}}$Sr$_{{\rm 1/8}}$MnO$_3$. Its phonon bending mode starts to be
splitted near a structural phase transition temperature. With further
cooling, strength of a mid-infrared small polaron peak, which is located at
0.4 eV, becomes increased. These temperature dependent changes in the phonon
spectra and the polaron absorption peaks were explained by a phase
separation and a percolation-type transition.
\end{abstract}



\newpage

There have been lots of attentions paid on physical properties of doped
manganites due to their scientific interests as well as potential
applications. For {\it x }$>$ 0.17, La$_{{\rm 1-}{\it x}}$Sr$_{{\it x}}$MnO$%
_3$ (LSMO) shows a metal-insulator (MI) transition and colossal
magnetoresistance (CMR) near a ferromagnetic (FM) ordering temperature, $T_C$%
. On the other hand, LSMO samples with 0.10 $\leq $ {\it x }$\leq $ 0.17
display a series of intriguing transitions. It experiences a structural
change at $T_S$ ($>T_C$). Just below $T_C$, its temperature-dependent
resistivity, $\rho $($T$), shows a metal-like behavior. However, it finally
enters into an insulating state at $T_P$ ($<T_C$).\cite{moritomo1}

Using neutron diffraction measurements, Kawano {\it et al.}\cite{kawano}
showed that the LSMO samples with 0.10 $\leq $ {\it x }$\leq $ 0.17 have two
kinds of orthorhombic phases: one is the $O^{\prime }$ phase with static
Jahn-Teller (JT) distortion, while the other is a pseudocubic phase, called
the ''$O^{*}$ phase''. In particular, they claimed that the {\it x}=1/8
sample exhibits $O^{*}\rightarrow O^{\prime }\rightarrow O^{*}$ structural
transitions at $T_S$ and $T_P$. From neutron scattering measurements, Yamada 
{\it et al.}\cite{yamada} observed appearance of satellite reflection peaks
below $T_P$. To explain this intriguing phenomenon, they used a
polaron/charge ordering model. They showed that, below $T_P$, undistorted
(001) planes where holes construct a 2$\times $2 square lattice alternate
with (MnO$_2$)$^{-}$ planes containing a static JT ordering of the oxygen
displacements as in the LaMnO$_3$ (001) planes. They also argued that the
locking into the commensurate structure should occur for LSMO samples within
a finite concentration range around the nominal value of {\it x}=1/8.

Later, Zhou {\it et al.}\cite{zhou} claimed that a dynamic segregation into
the hole-rich and the hole-poor regions should occur at $T_S$, even though
such a dynamic process would remain undetected by the diffraction
experiments. Moreover, in the temperature range of $T_P<T<T_C$, they argued
that the metallic behavior is not due to itinerant-electron conduction but
that it comes from three-manganese clusters.

Recently, a couple of groups suggested existence of a phase separation (PS)
in the lightly doped LSMO samples. Yunoki, Moreo, and Dagotto\cite{yunoki}
performed computational studies on the 2-orbital Kondo model with JT
phonons. With a moderate electron-phonon coupling, a PS was found to occur
between (i) a metallic spin-FM and orbital-disordered phase (caused by the
double-exchange interaction) and (ii) an insulating spin-FM and
orbital-antiferro (AF) ordered phase. Endoh {\it et al}.\cite{endoh}
observed an AF-type orbital ordering of {\it e}$_g$ electron below $T_P$ in
La$_{0.88}$Sr$_{0.12}$MnO$_3$ by an anomalous x-ray scattering experiment.
Using an effective Hamiltonian where spin and orbital degrees were treated
on an equal footing, they showed the PS between the spin-FM phases. However,
they claimed that the insulating phase was caused by superexchange
interaction between neighboring {\it e}$_g$ spins, so that the JT distortion
was irrelevant in the insulating phase.\cite{ahn}

In spite of these interesting physical phenomena, optical properties of La$%
_{7/8}$Sr$_{1/8}$MnO$_3$ have not been investigated in detail.\cite{CMR} In
this paper, we report details of our optical investigations, including $T$%
-dependent changes in phonon spectra and infrared (IR) absorption features.
These changes will be discussed in terms of a phase separation and a
percolation-type transition.

A polycrystalline La$_{7/8}$Sr$_{1/8}$MnO$_3$ sample was prepared by the
conventional solid-state reaction method.\cite{kimprl} X-ray powder
diffraction measurement showed that the sample did not contain secondary
phases, and electron-probe microanalysis indicated that its chemical
composition was stoichiometric within an error bar of 1 $\%$. The $T$%
-dependent resistivity was measured by the four-probe method and
magnetization, $M$, was determined using a SQUID magnetometer.

We measured reflectivity spectra $R$($\omega $) from 0.01 to 30 eV. To
obtain $T$-dependent $R$($\omega $) in the frequency region from 0.01 to 2.5
eV, we used a liquid He-cooled cryostat. Since there were less than 2 $\%$
changes near 2.5 eV, we attached the low temperature $R$($\omega $) smoothly
with room temperature data above the frequency. Optical conductivity spectra 
$\sigma $($\omega $) were obtained using the Kramers-Kronig (KK)
transformation. Details of reflectivity measurements and the KK analysis for
other manganite samples were described earlier.\cite{jung1} Spectroscopic
ellipsometry measurements were performed independently to obtain $T$%
-dependent $\sigma $($\omega $) in a frequency region between 1.5 and 5.0
eV. The optical conductivity data obtained from the spectroscopic
ellipsometry agreed quite well with those from the reflectivity
measurements, demonstrating validity of our KK analysis.

The inset of Fig. 1 shows $\rho $($T$) of La$_{7/8}$Sr$_{1/8}$MnO$_3$.
Measured values of $\rho $($T$) were quite close to the reported values.\cite
{pinsard} Around $T_S$ ($\sim $ 260 K), a weak thermal hysteresis behavior
was observed. In the paramagnetic state above $T_C$ ($\sim $ 190 K), La$%
_{7/8}$Sr$_{1/8}$MnO$_3$ shows an insulating behavior. Just below $T_C$, the 
$\rho $($T$) curve shows a metal-like behavior. However, below $T_P$ ($\sim $
160 K), it shows an upturn, indicating a reentrance into an insulating
state. At the low temperature region, $\rho $($T$) can be fitted reasonably
well with the variable range hopping model: $\rho $($T$)=$\rho _0$exp($\frac{%
\text{C}}{k_BT}$)$^{1/4}$.

Figure 1 shows $R$($\omega $) of La$_{7/8}$Sr$_{1/8}$MnO$_3$ at various
temperatures. The sharp peaks in the far-IR region are originated from optic
phonon modes. Around 0.03 and 0.06 eV, additional small peaks can be seen
clearly at low temperatures, indicating appearance of other phonon modes.
Even for the metal-like region of $T_P<T<T_C$, $R$($\omega $) do not show
any upturn behaviors at the low frequency region. [Reflectivity of a typical
metal shows a upturn behavior in the dc limit (i.e., $\omega \rightarrow $
0).] Below $T_C$, $R$($\omega $) increase sharply around 0.3 eV and decrease
slightly around 1.5 eV. In the temperature region of 200 K $\leq T\leq $ 300
K, $R$($\omega $) remain nearly the same. Above 2.5 eV, broad peaks due to
the interband transition between electronic levels can be observed.

Figure 2 displays $\sigma $($\omega $) of La$_{7/8}$Sr$_{1/8}$MnO$_3$ in the
far-IR region. There are three main optic phonon modes at 280 K: the phonons
around 160, 350, and 570 cm$^{-1}$ are known as external, bending, and
stretching modes, respectively.\cite{kimprl} Just below $T_S$, the bending
mode starts to be splitted slightly, and there are weak signatures for
additional phonon modes around 220, 280, and 490 cm$^{-1}$. The external and
the stretching modes are shifted to higher frequencies by about 3 cm$^{-1}$
without any apparent splitting. With further cooling, the bending mode
splitting and the additional phonon modes become more clear. The bending
mode splitting and the additional phonon modes seem to be related to the
polaron/charge ordering. With such an ordering, its lattice unit cell
becomes enlarged, so there should be changes in the phonon spectra. Similar
changes were observed below charge ordering temperature in La$_{1.67}$Sr$%
_{0.33}$NiO$_4$.\cite{katsufuji}

There are three important points which should be addressed in these phonon
spectra. First, the optical conductivity spectrum at 15 K ($\ll T_P$) is
quite different from that at 280 K ($>T_S$), even though both of the
crystallographic structures in these temperature regions are the pseudocubic 
$O^{*}${\it .}\cite{kawano} This demonstrates that the local symmetries in
these two phases should be different. Contrary to the claim by Endoh {\it et
al}.,\cite{endoh} the local JT distortion might persist below $T_P$, which
is consistent with recent pulsed neutron diffraction measurements.\cite
{louca} Second, the bending mode splitting starts to occur at $T_S$ (neither 
$T_P$ nor $T_C$), suggesting that the phonon spectra changes should be
closely related to the structural change. Third, there is little change in
the stretching phonon mode position. In the La$_{0.7}$Ca$_{0.3}$MnO$_3$,
which shows a MI\ transition and CMR near $T_C$, the stretching mode shows a
significant frequency shift of about 20 cm$^{-1}$.\cite{kimprl} This
suggests that the metal-like region of La$_{7/8}$Sr$_{1/8}$MnO$_3$ might
have characteristics different from those of CMR manganites below $T_C$.

The $T$-dependent $\sigma $($\omega $) in the mid-IR region are shown in
Fig. 3. The sharp peaks below 0.1 eV are due to optic phonons. Above $T_C$, $%
\sigma $($\omega $) show a broad peak around 1.5 eV. Below $T_C$, the
spectral weight (SW) around 0.4 eV increases significantly without apparent
changes in its peak position. In a small polaron picture, the position of
the mid-IR absorption should be about 4 times larger than corresponding
polaron activation energy.\cite{cox} If this mid-IR peak can be attributed
to the small polaron absorption, the activation energy in La$_{7/8}$Sr$%
_{1/8} $MnO$_3$ could be estimated to be about 0.1 eV, consistent with
recent transport measurements.\cite{zhao}

Note that the mid-IR features for La$_{7/8}$Sr$_{1/8}$MnO$_3$ are different
from those of other CMR manganites, including La$_{0.825}$Sr$_{0.175}$MnO$_3$%
,\cite{okimoto1} Nd$_{0.7}$Sr$_{0.3}$MnO$_3$,\cite{kaplan,quijada} and La$%
_{0.7}$Ca$_{0.3}$MnO$_3$.\cite{kim3} First, in the CMR manganites, the SW in
the high energy region is transferred to a Drude term, which is typically
small, and a broad mid-IR peak. And, the position of the mid-IR peak shows a
very strong $T$-dependence, which might be related to a crossover from small
to large polaron.\cite{kim3} However, the mid-IR peak position in La$_{7/8}$%
Sr$_{1/8}$MnO$_3$ is nearly independent of $T$. Second, even for the
metal-like region of $T_P<T<T_C$, $\sigma $($\omega $) of La$_{7/8}$Sr$_{1/8}
$MnO$_3$ do not show any Drude peak. This unusual behavior is consistent
with the idea by Zhou {\it et al.} that the sample remains in an
unconventional FM state without itinerant carriers.\cite{zhou} They argued
that the metallic behavior should be originated from three-manganese
clusters, in which the hole was alternately at one of the two Mn$^{3+}$ ions
and was made mobile by dynamic JT coupling to the oxygen vibrations between
manganese atoms.

The inset in Fig. 3 shows $\sigma $($\omega $) measured at 85 K and 250 K by
the spectroscopic ellipsometry in a photon energy region of 1.5 $\sim $ 4.0
eV: as $T$ decreases, $\sigma $($\omega $) decrease slightly with a broad
background. [A similar SW change in the visible-UV region was observed for a
La$_{0.6}$Sr$_{0.4}$MnO$_3$ thin film using a transmission measurement.\cite
{moritomo}] Within our experimental errors, the total change of the SW
measured by the spectroscopic ellipsometry is close to that observed below
1.5 eV, and with an opposite sign. This fact suggests that a significant
portion of the SW in the mid-IR region was transferred from those between
1.5 and 4.0 eV.

To investigate the SW changes quantitatively, we regarded $\sigma $($\omega $%
) as a sum of two components: $\sigma $($\omega $) $=\sigma _{ms}$($\omega $)%
$+\sigma _L$($\omega $), where $\sigma _{ms}$($\omega $) and $\sigma _L$($%
\omega $) represent contributions of midgap states and Lorentz oscillators,
respectively. $\sigma _L$($\omega $) mainly come from interband transition
of O\ 2{\it p} $\rightarrow $ {\it e}$_g$ and transitions from lower to
upper Hund's rule split bands, i.e. {\it e}$_g^{\uparrow }$({\it t}$%
_{2g}^{\uparrow }$) $\rightarrow $ {\it e}$_g^{\uparrow }$({\it t}$%
_{2g}^{\downarrow }$) and {\it e}$_g^{\downarrow }$({\it t}$%
_{2g}^{\downarrow }$) $\rightarrow $ {\it e}$_g^{\downarrow }$({\it t}$%
_{2g}^{\uparrow }$).\cite{moritomo,jung3} [This notation indicates that the
transitions occur between two {\it e}$_g$ bands with the same spin but under
different {\it t}$_{2g}$ spin backgrounds.] Following our earlier works,\cite
{jung3} we initially fitted $\sigma $($\omega $) above 2.0 eV with a series
of the Lorentz oscillator functions and then obtained $\sigma _{ms}$($\omega 
$) by subtracting $\sigma _L$($\omega $) from measured $\sigma $($\omega $).

The $T$-dependent $\sigma _{ms}$($\omega $) data are shown in Fig. 4, and
they were analyzed with double peak structures. The strong peak near 1.5 eV,
which will be called ''Peak II'', is nearly temperature independent. The
peak near 0.4 eV, which will be called ''Peak I'', is quite weak at 210 K,
but it becomes evident around 180 K (i.e. just below $T_C$). As $T$ becomes
lower, it becomes stronger. In our earlier doping dependent studies on
optical properties of La$_{{\rm 1-}{\it x}}$Ca$_{{\it x}}$MnO$_3$,\cite
{jung3} similar double peak structures were observed. And, recent
calculations by Yunoki, Moreo, and Dagotto\cite{yunoki} also predicted such
double peak structures in $\sigma $($\omega $). Based on these works, we
assigned Peaks I to a small polaron absorption.\cite{Peak_I} With this
assignment, the temperature dependence of Peak I can be explained
qualitatively. Below $T_C$, {\it t}$_{2g}$ spins are aligned to make it
easier for an {\it e}$_g$ electron to hop from a JT splitted Mn$^{3+}$ site
to an unsplitted Mn$^{4+}$ site.

A fitting procedure with two Gaussian functions was used to find more
quantitative behaviors of this double peak structure. Strengths of the
peaks, $S_{ms}^k$ ($k=$ I and II), were obtained by numerically integrating
the corresponding Gaussian functions. Values of $S_{ms}^{\text{I}}$ are
shown as the solid squares in Fig. 5. [Values of $S_{ms}^{\text{II}}$ are
nearly $T$-independent within our error bars.] Above $T_C$, $S_{ms}^{\text{I}%
}$ is small and nearly $T$-independent. Although the phonon spectra in Fig.
2 start to change at $T_S$, there is no apparent change in $S_{ms}^{\text{I}%
} $ near this temperature. With cooling, $S_{ms}^{\text{I}}$ sharply
increases below $T_C$ and becomes nearly saturated around 100 K. Although
the $T$-dependence of $S_{ms}^{\text{I}}$ is similar to that of $M/M_S$ ($%
\equiv M^{*}$), where $M_S$ is the saturated magnetization, there are a
couple of differences which should be noted. First, above $T_C$, $M$
vanishes, but $S_{ms}^{\text{I}}$ seems not. Second, $M$ saturates near $T_P$%
, but $S_{ms}^{\text{I}}$ saturates around 100 K.

Without polaron/charge ordering, $S_{ms}^{\text{I}}$ should depend simply on
a number of available polaron absorption transitions. Suppose that an {\it e}%
$_g$ electron at a JT splitted Mn$^{3+}$ ion has an up-spin (which is
parallel to $M$) and is located at a nearest neighbor site of a unsplitted Mn%
$^{4+}$ ion. Since the polaron hopping process should conserve the spin, its
strength should be proportional to $[1+M^{*}]^2/4$. For a down-spin, it
should be proportional to $[1-M^{*}]^2/4$. Therefore, $S_{ms}^{\text{I}}$
should be proportional to $[1+M^{*2}]/2$, which is the same as the behavior
of the Drude peak strength predicted for the CMR manganites in the
spin-split band picture.\cite{furukawa} However, the experimental values of $%
[1+M^{*}{}^2]/2$, shown as the solid line in Fig. 5, do not agree with $%
S_{ms}^{\text{I}}$. From our data, it is quite plausible to say that the
photon-assisted polaron transfer matrix might be enhanced significantly at
the polaron/charge ordered state, where $M$ is already nearly saturated.

One of possible scenarios to explain the $T$-dependence of phonon spectra
and $S_{ms}^{\text{I}}$ is a percolation-type transition.\cite{two_fluid} As
suggested by Zhou {\it et al.},\cite{zhou} a dynamic phase segregation into
the hole-rich and the hole-poor regions (i.e. formation of the
polaron/charge ordered domains) might start to occur around $T_S$. Changes
in the phonon mode will appear, and $S_{ms}^{\text{I}}$ remains finite but
small. At the temperature region of $T_P<T<T_C$, the FM ordering starts to
occur and $S_{ms}^{\text{I}}$ increase very rapidly. In this temperature
region, a PS might occur between two spin-FM phases, i.e. a metallic
orbital-disordered phase and an insulating orbital-ordered phase, as
suggested by Yunoki {\it et al.}\cite{yunoki} and by Endoh {\it et al}.\cite
{endoh} Near $T_C$, the metallic phase is favored because entropy promotes
the orbital disordering and the carrier mobilities. However, below $T_P$,
the insulating FM orbital-ordered phase becomes dominant, resulting the
polaron/charge ordering observed by Yamada {\it et al}.\cite{yamada} $%
S_{ms}^{\text{I}}$ will increase until all of the sample becomes insulating
in the polaron/charge ordered state. Even though this picture based on the
PS and a percolation-type transition can explain most $T$-dependent changes
in $\sigma $($\omega $) and $\rho $($T$), more studies are required to get a
more clear picture.

In summary, we investigated optical conductivity spectra of La$_{7/8}$Sr$%
_{1/8}$MnO$_3$. The phonon bending mode starts to be splitted near a
structural transition temperature. At lower temperature, the strength of the
mid-infrared small polaron peak becomes increased. Even in the metal-like
region, the Drude peak was not observed. These features supports that a
phase separation and a percolation-type transition should occur in La$_{7/8}$%
Sr$_{1/8}$MnO$_3$.

We acknowledge valuable discussions with Prof. E. J. Choi, Prof. Jaejun Yu,
and K. H. Ahn. This work are financial supported by the Ministry of
Education through the Inter-University Center for Natural Science Research
Facilities (BSRI-97-7401), by the Korea Science and Engineering Foundation
through grant No. 971--0207-024-2 and through RCDAMP of Pusan National
University. Experiments at PLS were supported in part by MOST and POSCO. 

\begin{figure}[tbp]
\caption{Temperature dependent $R$($\omega $) of La$_{7/8}$Sr$_{1/8}$MnO$_3$%
. In the inset, $\rho $($T$) for the same sample is shown. }
\label{Fig:1}
\end{figure}

\begin{figure}[tbp]
\caption{Optic phonon modes of La$_{7/8}$Sr$_{1/8}$MnO$_3$. Positions of the
external and the stretching phonon frequencies at 280 K are drawn as dotted
lines.}
\label{Fig:2}
\end{figure}

\begin{figure}[tbp]
\caption{Temperature dependent $\sigma $($\omega $) of La$_{7/8}$Sr$_{1/8}$%
MnO$_3$ below 2.0 eV. In the inset, $\sigma $($\omega $) obtained from the
spectroscopic ellipsometry are shown.}
\label{Fig:3}
\end{figure}

\begin{figure}[tbp]
\caption{The midgap state conductivity $\sigma _{ms}$($\omega $) of La$%
_{7/8} $Sr$_{1/8}$MnO$_3$. The solid circles and the dotted lines represent
the experimental data and the Gaussian fitting lines, respectively.}
\label{Fig:4}
\end{figure}

\begin{figure}[tbp]
\caption{Values of $S_{ms}^{\text{I}}$ in La$_{7/8}$Sr$_{1/8}$MnO$_3$.
Predicted values of $M^{*}$, $M^{*2}$, and $[1+M^{*2}]/2$ were denoted as
the dotted, the dot-dashed, and the solid lines, respectively. }
\label{Fig:5}
\end{figure}

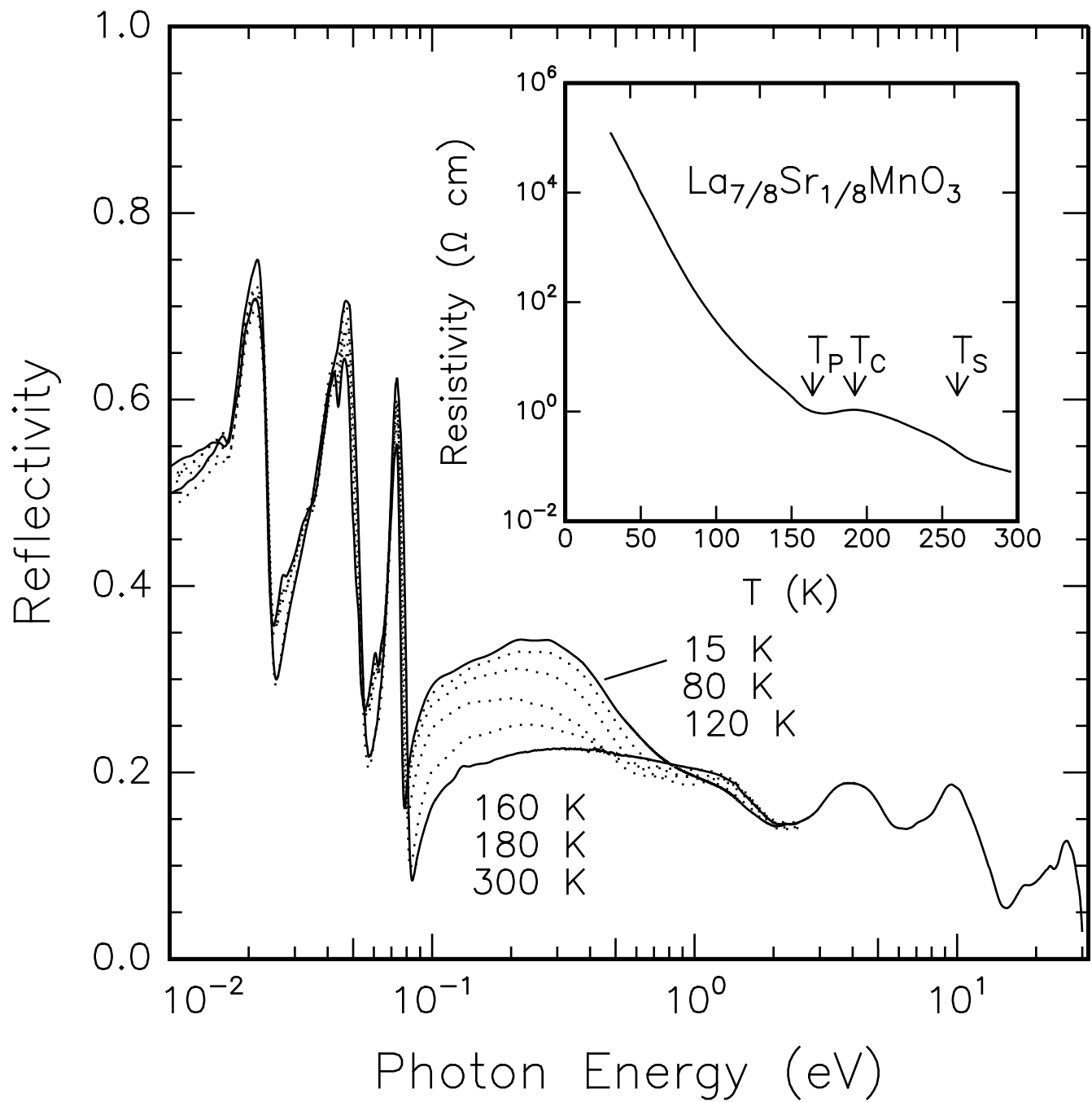
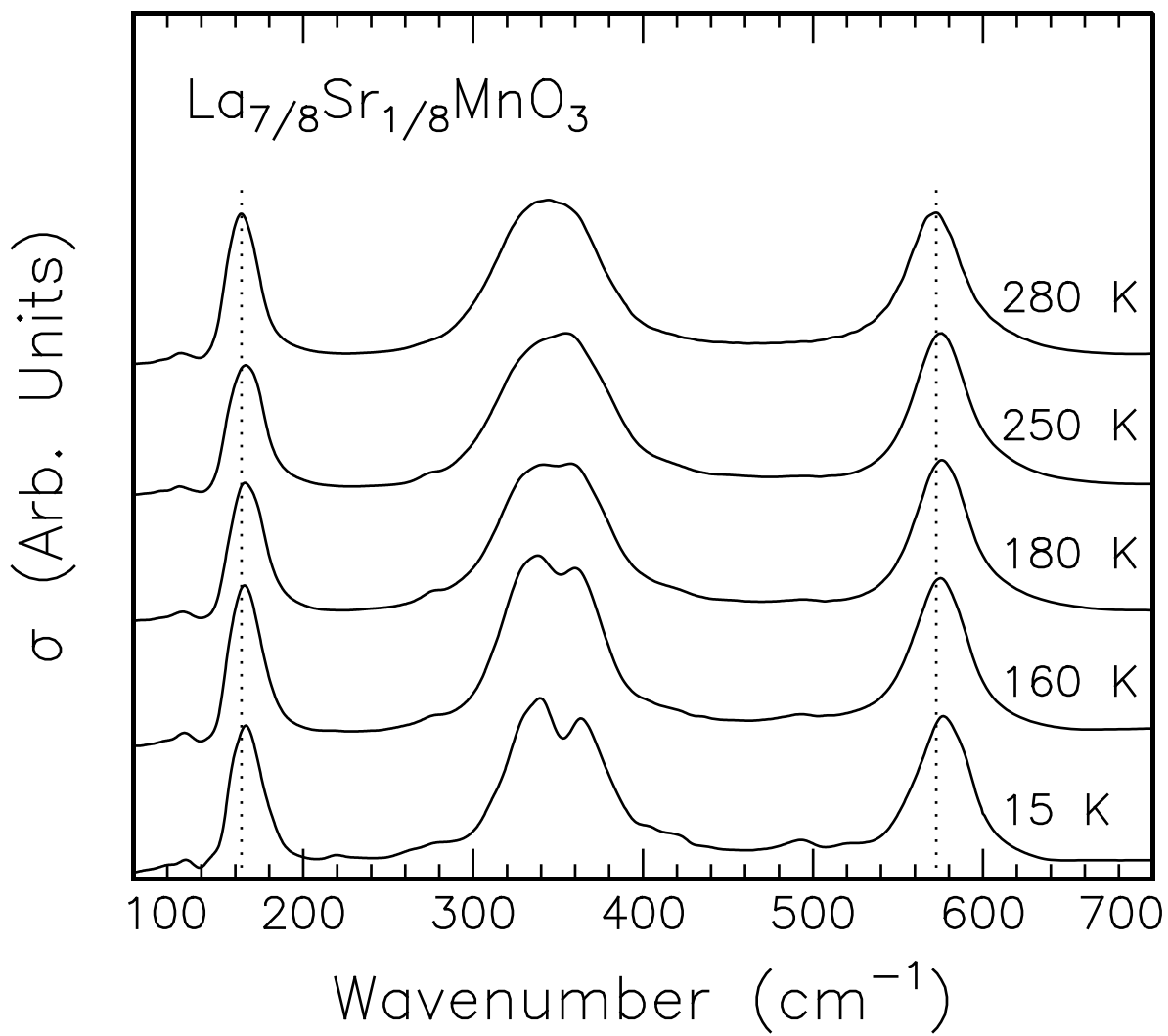
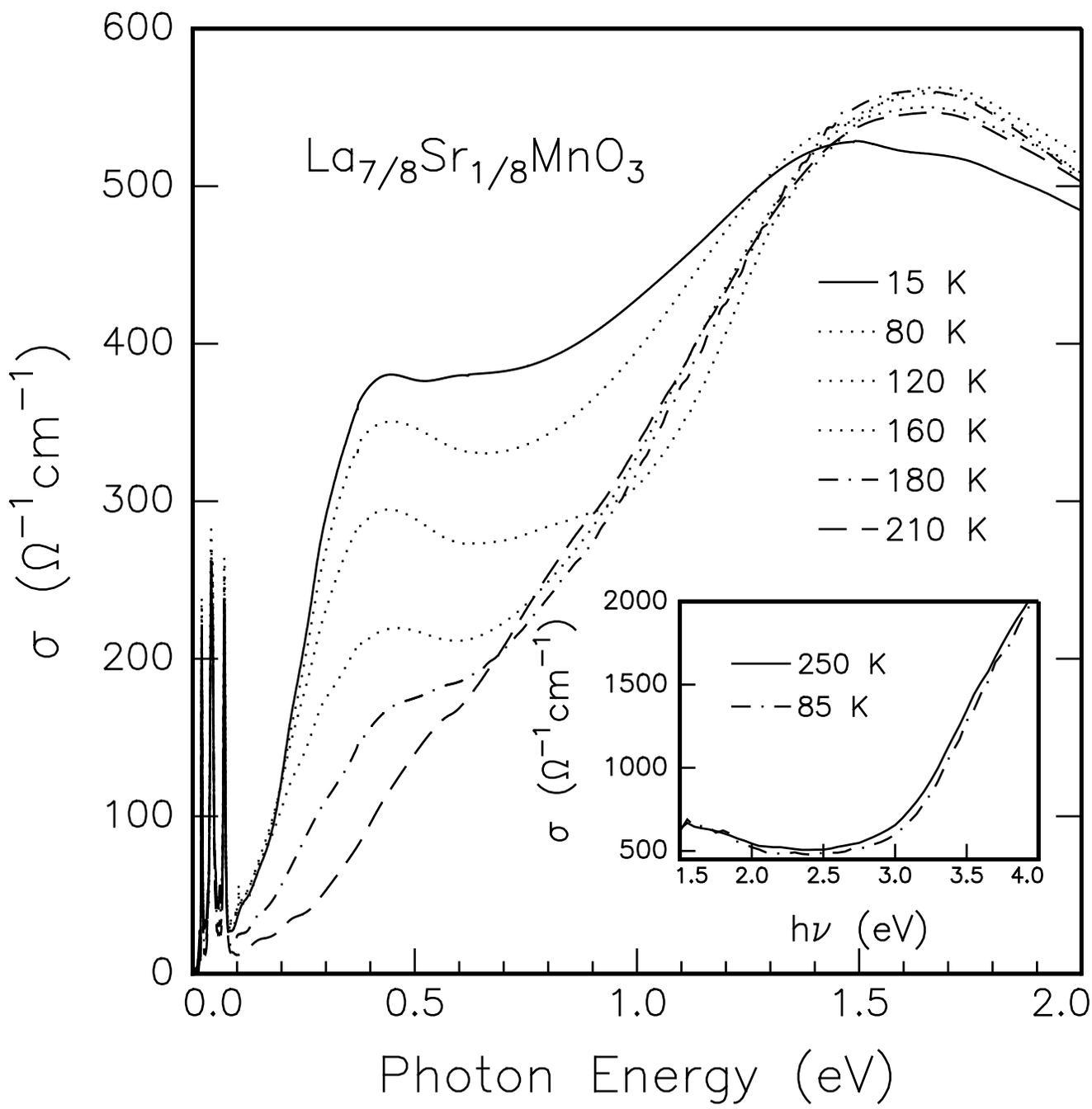
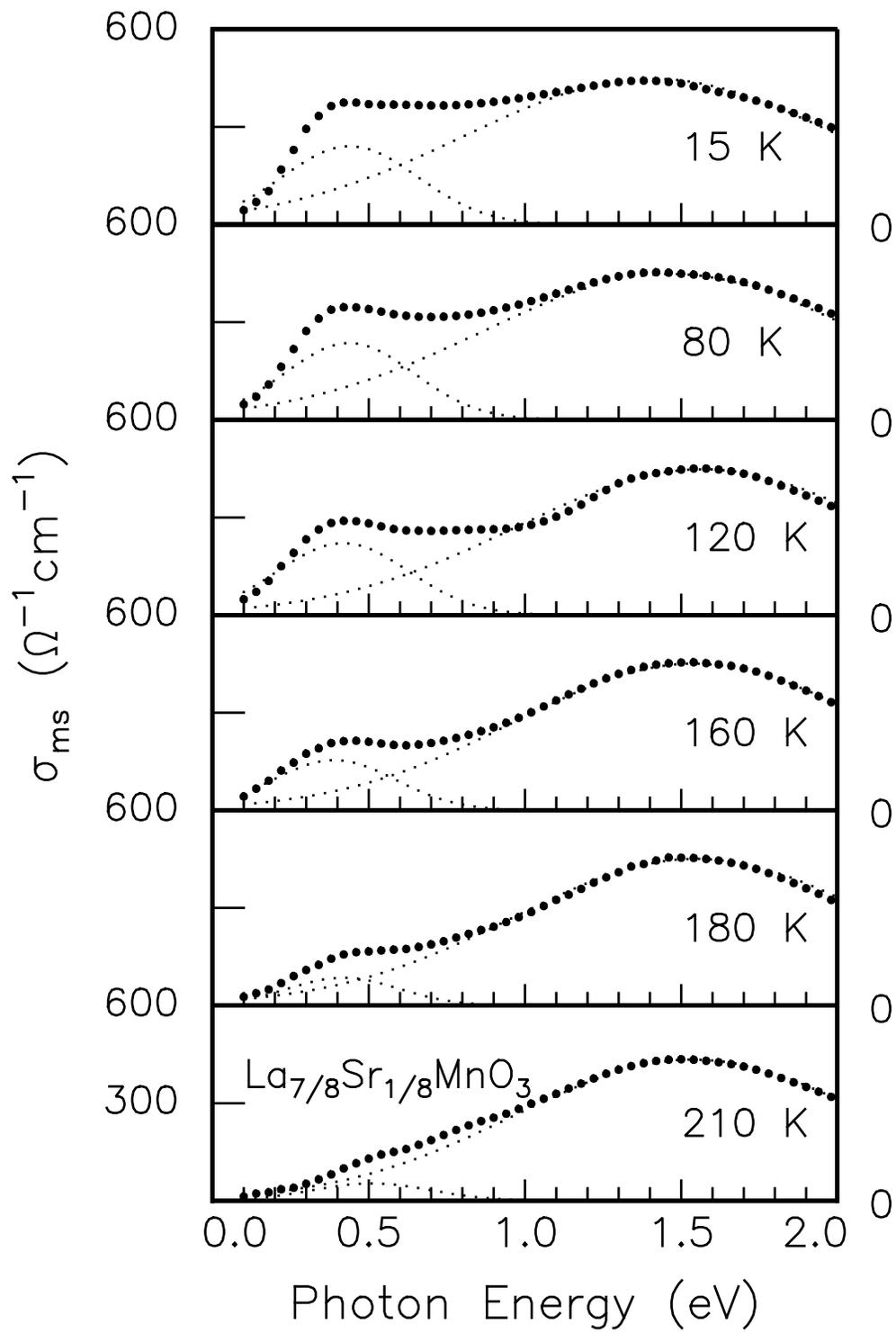
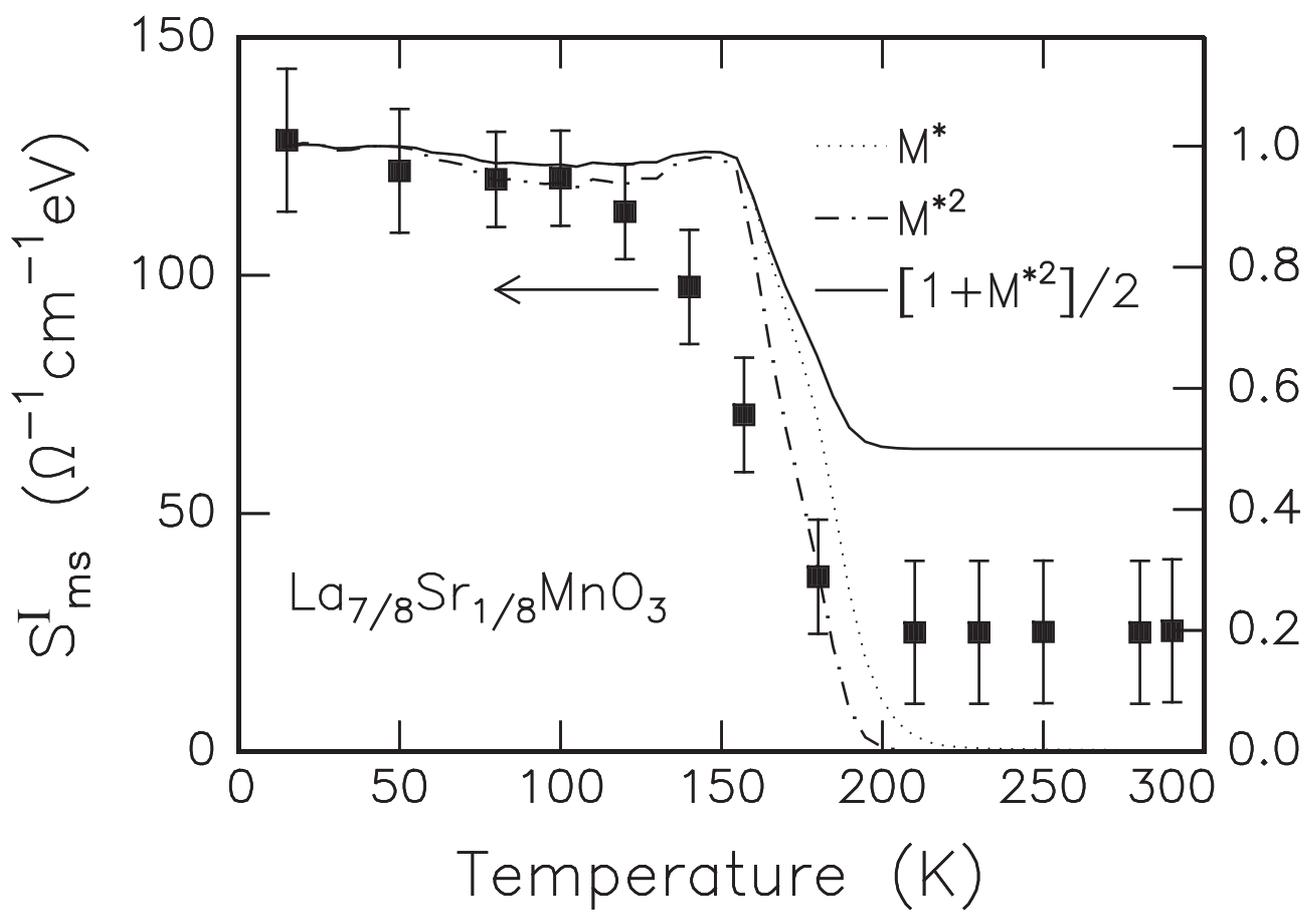
\end{document}